\documentclass[usenatbib,a4paper,usenames,dvipsnames]{mn2e}
\usepackage{amsmath}
\usepackage{amssymb}
\usepackage{color,hyperref}
\definecolor{darkblue}{rgb}{0.0,0.0,0.3}
\hypersetup{colorlinks,breaklinks,
          linkcolor=darkblue,urlcolor=darkblue,
          anchorcolor=darkblue,citecolor=darkblue}
\usepackage{txfonts}

\DeclareSymbolFont{cmletters}{OML}{cmm}{m}{it}
\DeclareMathSymbol{v}{\mathalpha}{cmletters}{"76}
\usepackage{graphicx}

\newcommand{\useiop}{-3}

\newcommand{\shortauthors}[1]{}
\newcommand{\shorttitle}[1]{}
\newcommand{\altaffiltext}[2]{}
\newcommand\apj{\rmfamily{ApJ}}%
\newcommand\apjl{\rmfamily{ApJ}}%
\newcommand\apjs{\rmfamily{ApJS}}%
\newcommand\aap{\rmfamily{A\&A}}%
\newcommand\mnras{\rmfamily{MNRAS}}%
\newcommand\prd{\rmfamily{Phys.~Rev.~D}}%
\newcommand\jcap{\rmfamily{J.~Cosm.~Astrop.~Phys.}}%
\usepackage{ifthen}

\defcitealias{tch08}{TMN08}
\defcitealias{tch09}{TMN09}
\defcitealias{tch11}{TNM11}
\defcitealias{kom09}{K09}
\defcitealias{bz77}{BZ}
\defcitealias{bp82}{BP}
\defcitealias{spit06}{S06}
\defcitealias{kom06}{K06}

\newcommand{\const}{{\rm constant}}

\newcommand{\myvec}{\vec}

\newcommand{\cut}[1]{\hbox{}}
\newcommand{\cmt}[1]{}

\newcommand{\Fermi}{\emph{Fermi}}

\newcommand{\harm}{\texttt{HARM}}
\newcommand{\epsfive}{\epsilon}

\shortauthors{}
\shorttitle{}

\ifthenelse{\equal{\useiop}{-3}}{ 

\author[A.~Tchekhovskoy,
A.~Spitkovsky, and J.G.~Li]
{Alexander Tchekhovskoy$^1$\thanks{\hbox{E-mail:
      atchekho@princeton.edu~(AT)}},
Anatoly Spitkovsky$^2$, and
Jason G.~Li$^2$
\\
  $^1$Center for Theoretical Science, Jadwin Hall, Princeton University, Princeton,
  NJ 08544, USA; Princeton Center for Theoretical Science Fellow \\
 $^2$Department of Astrophysical Sciences, Peyton Hall, Princeton University, Princeton, NJ 08544, USA}
}{
}
\begin{document}
\label{firstpage}

\title[Oblique 3D MHD Pulsar Magnetospheres]%
{Time-Dependent 3D Magnetohydrodynamic Pulsar Magnetospheres:
Oblique Rotators}

\ifthenelse{\equal{\useiop}{-2}}{
\journal{New Astronomy}
\begin{frontmatter}

\author[cfa]{Alexander Tchekhovskoy}
\ead{atchekho@princeton.edu}

\address[cfa]{Princeton Center for Theoretical Science, Jadwin Hall, Princeton
  University, Princeton, NJ 08544, USA}
}{}
\ifthenelse{\equal{\useiop}{3}}{
\author{Alexander Tchekhovskoy$^1$ and $^2$} 
  \maketitle
  \begin{affiliations}
    \item Princeton Center for Theoretical Science, Princeton
  University, Jadwin Hall, Princeton
  NJ 08544; atchekho@princeton.edu\\
  \end{affiliations}
}{}
\ifthenelse{\equal{\useiop}{0}}{self-consistent 
\altaffiltext{1}{Princeton Center for Theoretical Science, Princeton
  University, Jadwin Hall, Princeton
  NJ 08544; atchekho@princeton.edu}
}

\ifthenelse{\equal{\useiop}{-3}}{ 
\date{Accepted . Received ; in original form }
\pagerange{\pageref{firstpage}--\pageref{lastpage}} \pubyear{2012}
\maketitle
}

\begin{abstract}
  The current state of the art in pulsar magnetosphere modeling
  assumes the force-free limit of magnetospheric plasma. This limit
  retains only partial information about plasma velocity and neglects
  plasma inertia and temperature.  We carried out time-dependent 3D
  relativistic magnetohydrodynamic (MHD) simulations of oblique pulsar
  magnetospheres that improve upon force-free by retaining the full
  plasma velocity information and capturing plasma heating in strong
  current layers.  We find rather low levels of magnetospheric
  dissipation, \hbox{with less than} $10$\% of pulsar spindown energy
  dissipated within a few light cylinder radii, and the MHD spindown
  that is consistent with that in force-free. While oblique
  magnetospheres are qualitatively similar to the rotating
  split-monopole force-free solution at large radii, we find
  substantial quantitative differences with the split-monopole, e.g.,
  the luminosity of the pulsar wind is more equatorially concentrated
  than the split-monopole at high obliquities, and the flow velocity
  is modified by the emergence of reconnection flow directed into the
  current sheet.
\end{abstract}

\ifthenelse{\equal{\useiop}{-2}}{
\end{frontmatter}
}{}

\ifthenelse{\equal{\useiop}{-3}}{ 
}{}
\ifthenelse{\equal{\useiop}{3}}{
  }{}
  
\ifthenelse{\equal{\useiop}{0}}{
{
  }
}


\begin{figure*}
  \begin{center}
    \includegraphics[width=0.32\textwidth]{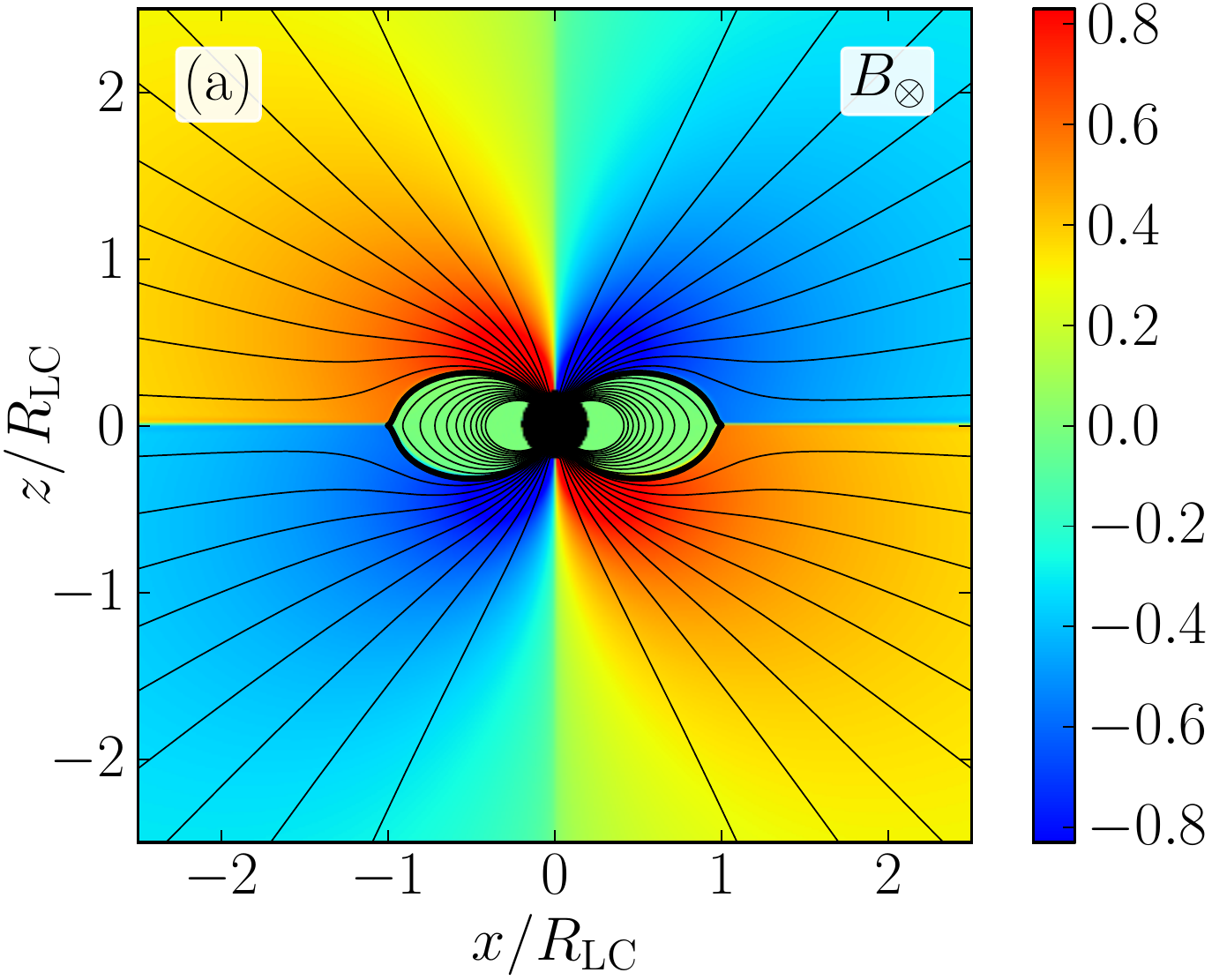}\hfill%
    \includegraphics[width=0.32\textwidth]{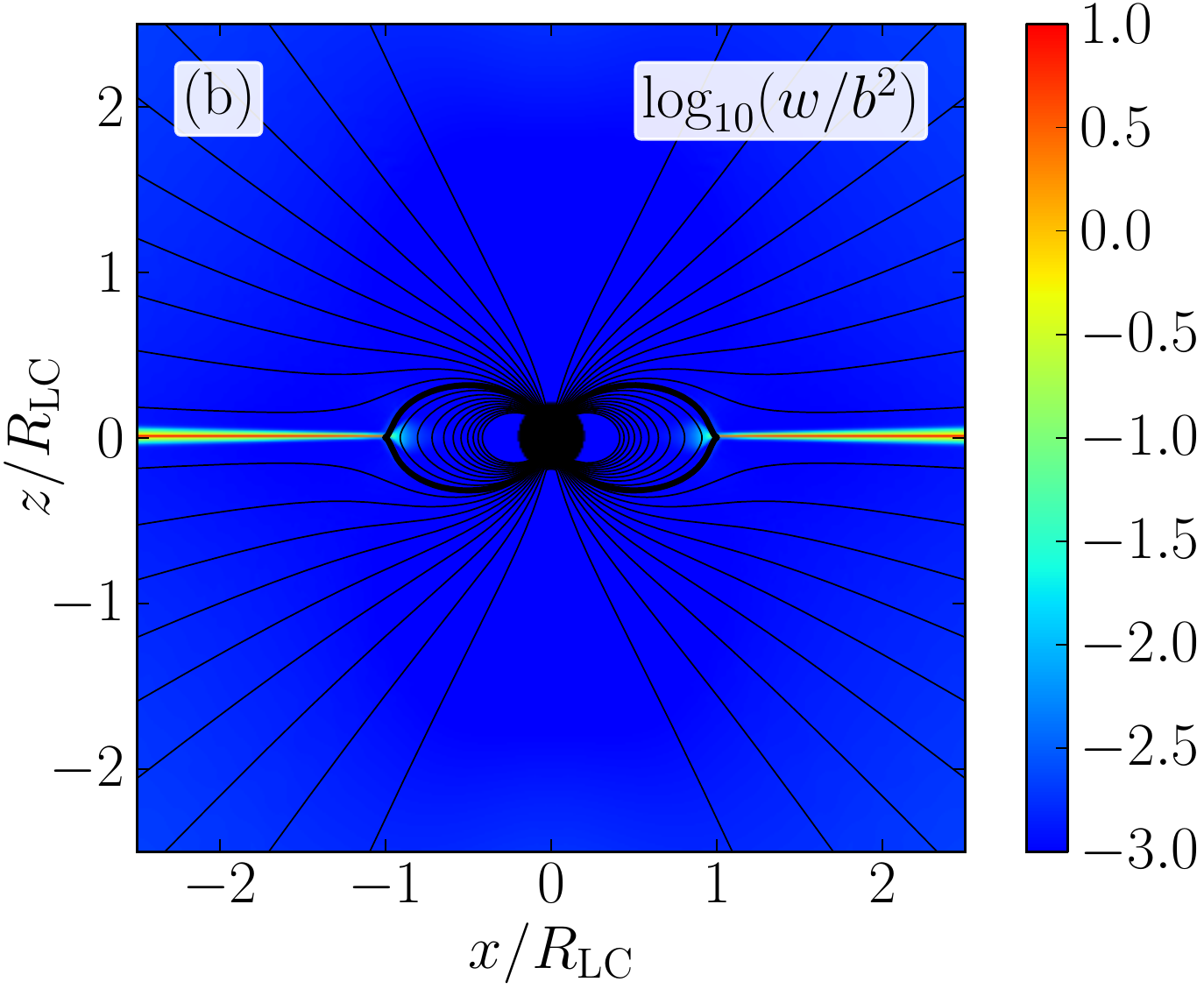}\hfill%
    \includegraphics[width=0.33\textwidth]{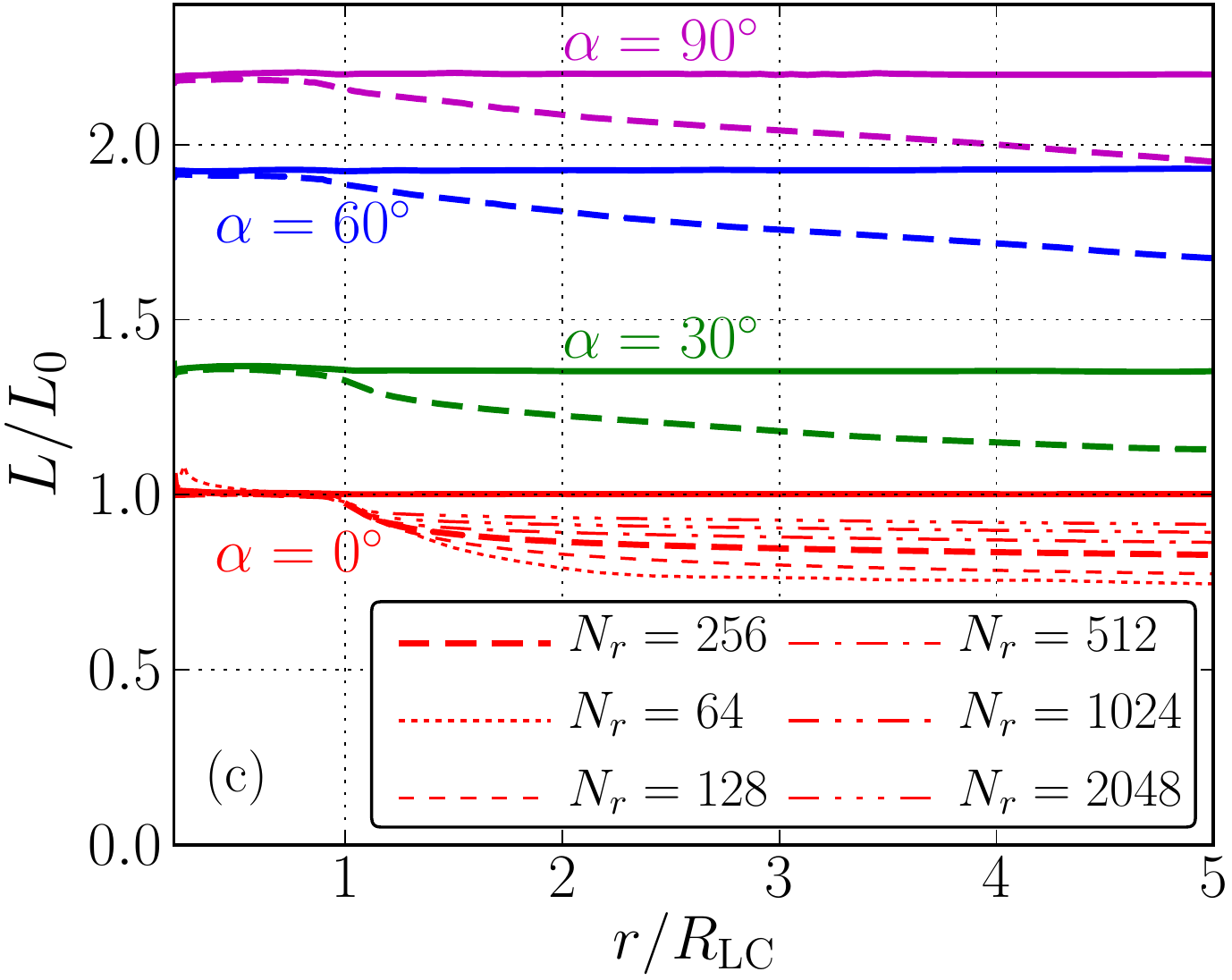}
  \end{center}
\vspace{-0.3cm}
  \caption{{\bf [Panels (a) and (b)]} 
    Meridional cut through a relativistic MHD
    magnetosphere of an aligned pulsar (model D0R2048) taken after $20$
    rotations;  by this time the shown part of the magnetosphere reached
    steady state.  Solid lines show contours of enclosed magnetic flux
    which represent poloidal field lines. 
    {\bf [Panel (a)]}
    Colour shows out-of-plane magnetic field component, $B_\otimes$
    (red/blue -- pointing into/out of plane, see colour bar),
    associated with poloidal currents circulating in the
    magnetosphere.  
    The boundary of the closed zone is shown with thick solid line.  
    {\bf [Panel (b)]} Colour shows the logarithm of the ratio of enthalpy, $w=\rho+\Gamma u_g$, to
    $b^2$ (red shows high and blue low values, see colour bar): $w/b^2$ is small near the star, where the magnetosphere is highly
    magnetized ($w/b^2\lesssim10^{-3}$, the magnetization is about $10\times$
    that in \citetalias{kom06}). However, $w/b^2$ exceeds unity
    in the equatorial current sheet, in which thermal pressure slows
    down plasma inflow and affects the velocity
    structure inside and near the sheet. {\bf [Panel (c)]} Radial
    profiles of angle-integrated energy flux, $L$,
    normalized to $L_0$, the analytic
  approximation for the spin-down of an aligned rotator, see
  eq.~\eqref{eq:alignededot}:  solid (dashed) lines show total
  (Poynting) energy flux. From bottom to top, models D0R64--D0R2048 (red),
  D30 (green), D60 (blue), D90 (magenta lines) are shown.  For
  an aligned pulsar, Poynting \hbox{fluxes at different resolutions are
  shown.} 
  }
\label{fig:aligned}
\end{figure*}

\section{Introduction}
\label{sec:introduction}

The launch of the \Fermi\ satellite opened a new window into studying
the properties of pulsar magnetospheres, 
with more than $100$ $\gamma$-ray
pulsars detected to date \citep{nolan12}. Since $\gamma$-ray emission can comprise a substantial
($\lesssim 10\%$) fraction of total pulsar spin-down losses,
$\gamma$-ray production mechanism 
should be capable of efficiently converting magnetospheric electromagnetic energy
into $\gamma$-ray radiation.  
Historically, much effort went into studying vacuum pulsar
magnetospheres based on the analytic magnetic field solution by \citet{deutsch55}, even though it was realized early on that for
pulsars, which are filled with abundant plasma  \citep{gol69}, force-free
approximation is a more appropriate framework. 
Force-free approximation accounts for magnetospheric charges and
currents but neglects plasma inertia, which is appropriate in the limit of high magnetization of the plasma $\sigma \equiv b^2/4 \pi \rho c^2 \gg 1$, with $b$ and $\rho$ the proper magnetic field and density. 
In recent years, self-consistent force-free solutions of axisymmetric
\citep*{ckf99,gruzinov_pulsar_2005,mck06pulff,tim06,par12phaedra} and oblique
(\citealt{spit06}, \citetalias{spit06} hereafter; \citealt{kc09,lst11,petri12a}) pulsar magnetospheres were numerically
obtained. They feature a magnetospheric current sheet that could
 be responsible for powering the observed $\gamma$-ray emission
\citep[e.g.,][]{bs09b,bs09a,petri12b,2012arXiv1208.2819A,2012arXiv1210.3346U}.

Magnetohydrodynamic (MHD) approach has important advantages relative to
force-free because it 
includes plasma inertia, pressure, and 
 velocity
component along the magnetic field, all of which are missing in the force-free model.
This information is crucial for modeling current sheet physics and for constructing 
realistic $\gamma$-ray light curves. Thus MHD holds the key to understanding global
structure, dissipation, and emission processes in pulsar magnetospheres.
The first 2D MHD simulation of the aligned pulsar magnetosphere was presented by
\citet{kom06}, \citetalias{kom06} hereafter.  
In this Letter we extend this work and construct time-dependent
3D relativistic MHD models of both axisymmetric and oblique pulsar
magnetospheres and make comparisons to force-free models.
In \S\ref{sec:numer-meth-probl} we describe our numerical method and
problem setup, and in \S\ref{sec:results} we present our results on
magnetospheric structure and
compare it to the split-monopole wind that is often
used to describe asymptotic structure of oblique pulsar 
magnetospheres.

\section{Numerical Method and Problem Setup}
\label{sec:numer-meth-probl}

\begin{table}
\begin{center}
\caption{Model details: model name, pulsar obliquity angle ($\alpha$),
  simulation resolution (given as $N_r\times N_\theta \times
  N_\varphi$), final time of the simulation ($t_f$, measured in units
  of pulsar period, $P$), pulsar spindown luminosity ($L$, measured in
units of aligned pulsar spindown luminosity, $L_{0}$,
eq.~\ref{eq:alignededot}), fraction of spindown luminosity dissipated
in the inner wind zone, $r<5R_{\rm LC}$ ($\epsilon$).}
\begin{minipage}{\columnwidth}
\begin{center}
  \begin{tabular}{l@{$\qquad$}r@{$\quad$}c@{$\quad$}c@{$\qquad$}l@{$\quad$}c@{$\quad$}}
\hline
Name &  $\alpha\ [^\circ]$  & Resolution & $t_f/P$ & $L/L_{\rm
  aligned}$ & $\epsilon\ [\%]$\\
\hline 
\multicolumn{6}{|c|}{Relativistic MHD models (with \harm{}):} \\
   D0R64 & $0$ & $64\times32\times1$ & $ 33$ & $0.929$ & $ 28$  \\ 
  D0R128 & $0$ & $128\times64\times1$ & $ 45$ & $0.983$ & $ 23$  \\ 
      D0 & $0$ & $256\times128\times1$ & $ 63$ & $  1$ & $ 17$  \\ 
  D0R512 & $0$ & $512\times256\times1$ & $ 44$ & $  1$ & $ 14$  \\ 
 D0R1024 & $0$ & $1024\times512\times1$ & $ 45$ & $0.994$ & $ 11$  \\ 
 D0R2048 & $0$ & $2048\times1024\times1$ & $ 22$ & $0.988$ & $8.7$  \\ 
     D15 & $15$ & $256\times128\times128$ & $5.2$ & $1.13$ & $ 19$  \\ 
     D30 & $30$ & $256\times128\times128$ & $5.1$ & $1.36$ & $ 17$  \\ 
     D45 & $45$ & $256\times128\times128$ & $5.1$ & $1.64$ & $ 15$  \\ 
  D60R64 & $60$ & $64\times32\times32$ & $4.8$ & $1.92$ & $ 27$  \\ 
 D60R128 & $60$ & $128\times64\times64$ & $5.5$ & $1.94$ & $ 16$  \\ 
     D60 & $60$ & $256\times128\times128$ & $3.3$ & $1.92$ & $ 13$  \\ 
 D60R512 & $60$ & $512\times256\times256$ & $3.5$ & $1.96$ & $ 12$  \\ 
     D75 & $75$ & $256\times128\times128$ & $2.9$ & $2.14$ & $ 12$  \\ 
     D90 & $90$ & $256\times128\times128$ & $  5$ & $2.2$ & $ 11$  \\ 

\multicolumn{6}{|c|}{Force-free models (with \harm{}):} \\
 D0R64ff & $0$ & $64\times32\times1$ & $7.7$ & $0.839$ & $ 50$  \\ 
D0R128ff & $0$ & $128\times64\times1$ & $ 30$ & $0.886$ & $ 47$  \\ 
    D0ff & $0$ & $256\times128\times1$ & $240$ & $0.914$ & $ 45$  \\ 
D0R512ff & $0$ & $512\times256\times1$ & $ 10$ & $0.925$ & $ 45$  \\ 
D0R1024ff & $0$ & $1024\times512\times1$ & $ 16$ & $0.93$ & $ 43$  \\ 
D0R2048ff & $0$ & $2048\times1024\times1$ & $ 13$ & $0.932$ & $ 43$  \\ 
   D30ff & $30$ & $256\times128\times128$ & $6.9$ & $1.27$ & $ 23$  \\ 
   D60ff & $60$ & $256\times128\times128$ & $  7$ & $1.83$ & $5.8$  \\ 
   D90ff & $90$ & $256\times128\times128$ & $  7$ & $2.11$ & $  2$  \\ 

\hline
\label{tab:models}
\end{tabular}
\end{center}
\end{minipage}
\vspace{-1cm}
\end{center}
\end{table}

We use general relativistic MHD code \harm\ \citep{gam03,mck04}
including recent improvements \citep{tch07,tch11,mb09,mtb12}.
For direct comparison to
previous studies, we neglect stellar gravity and carry out the
simulations in flat space.  We place the surface of the neutron star
(NS) at $r_* = 0.2 R_{\rm LC}$, where $R_{\rm LC} = c/\Omega$ is light
cylinder (LC) radius, $\Omega$ is pulsar angular frequency, and
$P=2\pi/\Omega$ is the pulsar period.  
We use a spherical
polar computational grid, $r$, $\theta$, $\varphi$, with $\theta=0$
along the rotation axis, $\myvec\Omega$. We also make use of cylindrical
radius, $R=r\sin\theta$. The grid extends
from the NS surface, $r_{\rm in} = r_* = 0.2 R_{\rm LC}$, to $r_{\rm
  out} = 200R_{\rm
  LC}$. 
Grid spacing is uniform in $\theta-$ and $\varphi-$directions. 
The spacing is logarithmic in the radial direction, $\Delta r/r=\const$, at
$r\lesssim r_0=20R_{\rm LC}$, and becomes progressively sparse,
$\Delta r/r \propto (\log r)^{3/4}$, at $r\gtrsim
r_0$.

We initialize the simulations with a dipolar magnetic field of 
a dipole moment, $\myvec\mu$, that makes
an angle, $\alpha$, with the rotation axis, $\myvec\Omega$. At
the inner $r-$boundary (the stellar surface), we set the
perpendicular (to the magnetic field) 3-velocity component, $\myvec
v_\perp$, to enforce stellar rotation, and set the parallel 3-velocity
component to zero, $\myvec v_{||}=0$.
We apply at
the outer $r-$boundary standard outflow boundary conditions (BCs), at
$\theta-$boundaries  transmissive polar BCs
\citep{mtb12},  and at $\varphi$-boundaries periodic BCs.

In a typical pulsar the magnetization near the LC can be very high, $\sigma_{\rm
  LC}\sim10^4$.  In a  
dipolar field, quantities drop off rapidly with $r$: $\rho\propto
b\propto \sigma \propto r^{-3}$. This is numerically challenging: to ensure force-free-like
conditions at the LC, $\sigma_{\rm LC}\gg1$, we must have a
very high $\sigma$ near the star, $\sigma_* = (R_{\rm
  LC}/R_*)^3\sigma_{\rm LC} \simeq 10^2 \sigma_{\rm LC}$, i.e.,  
much
higher than $\sigma\sim 10^2$ that our code can handle at reasonable
resolutions in 3D.
For our simulations to closely resemble the actual pulsar
magnetospheres, we ensure force-free-like
conditions inside the LC by driving density, $\rho$, internal energy,
$u_g$, and the spatial part of parallel 4-velocity, $u_{||}\equiv\gamma v_{||}\equiv(\myvec B\cdot
\myvec u)\,{\rm sign}(B_r)/B^2$, to target values, $\rho_t = b^2/4\pi c^2\sigma_t$
(with $\sigma_t=50$ or $100$), $u_{g,t} = 0.2\rho_t c^2$, and
$u_{||,t}=0$. We do this by modifying the variables, $q=\{\rho,u_g,u_{||}\}$, at
the end of each time step via $q = q_t +(q-q_t)\exp(-\kappa \Delta
t/\tau)$, where $\kappa=\{\cos\theta_m,\cos\theta_m,1\}$,
$\theta_m$ is the magnetic colatitude,
$\Delta t$ is the time step. For the driving timescale, we choose $\tau = 0$ at $r\le r_1=0.5R_{\rm LC}$, $\tau =
\tau_0 (r-r_1)(r_2-r_*)/[(r_2-r_1)(r_2-r)]$ at $r_1<r<r_2=R_{\rm LC}$,
and $\tau=\infty$ at $r\ge r_2$, where
$\tau_0=0.005P$ (see also \citetalias{kom06}).

\section{Results}
\label{sec:results}

\begin{figure*}
  \begin{center}
    \includegraphics[width=0.32\textwidth]{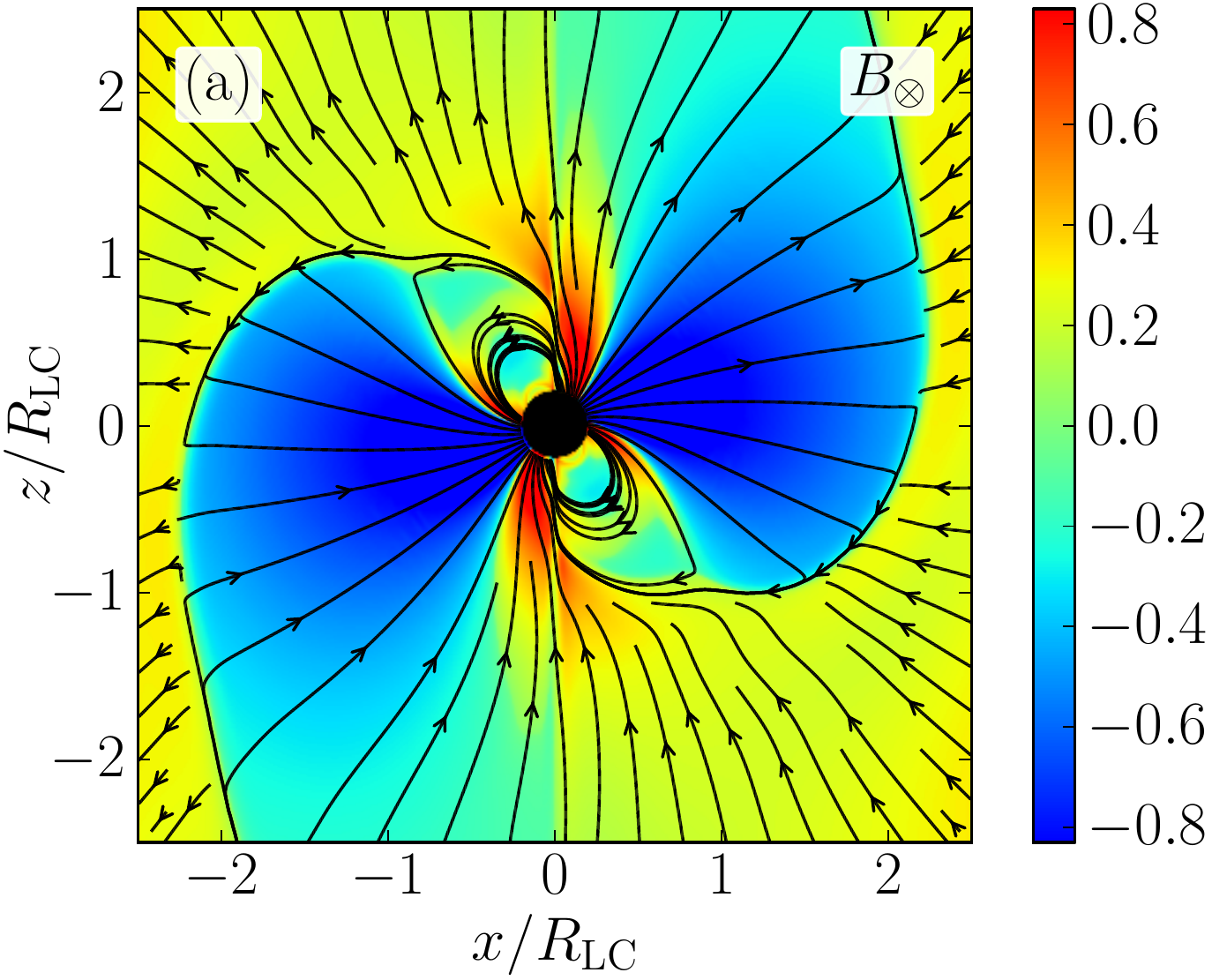}\hfill\
    \includegraphics[width=0.32\textwidth]{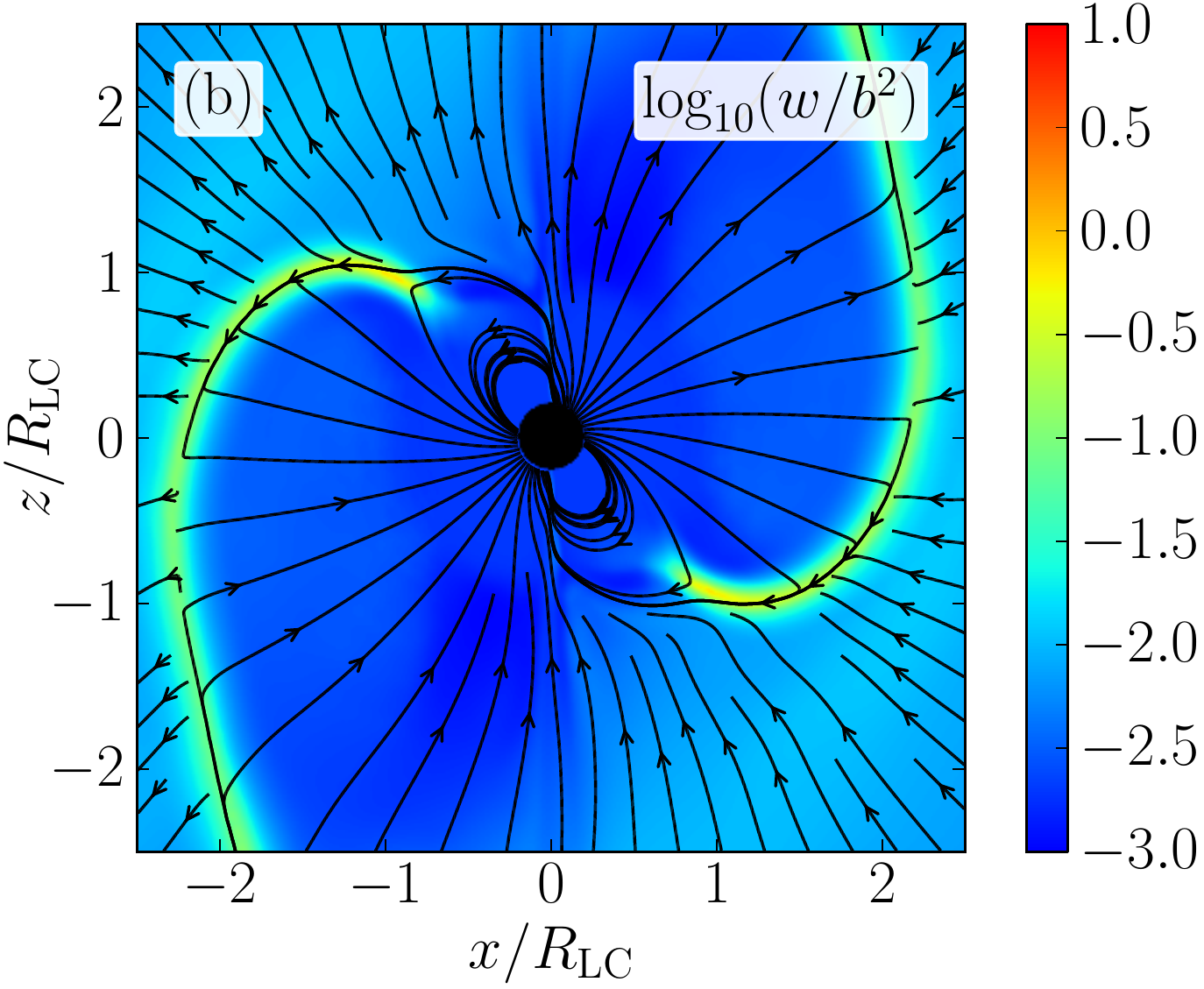}\hfill\
    \includegraphics[width=0.3\textwidth]{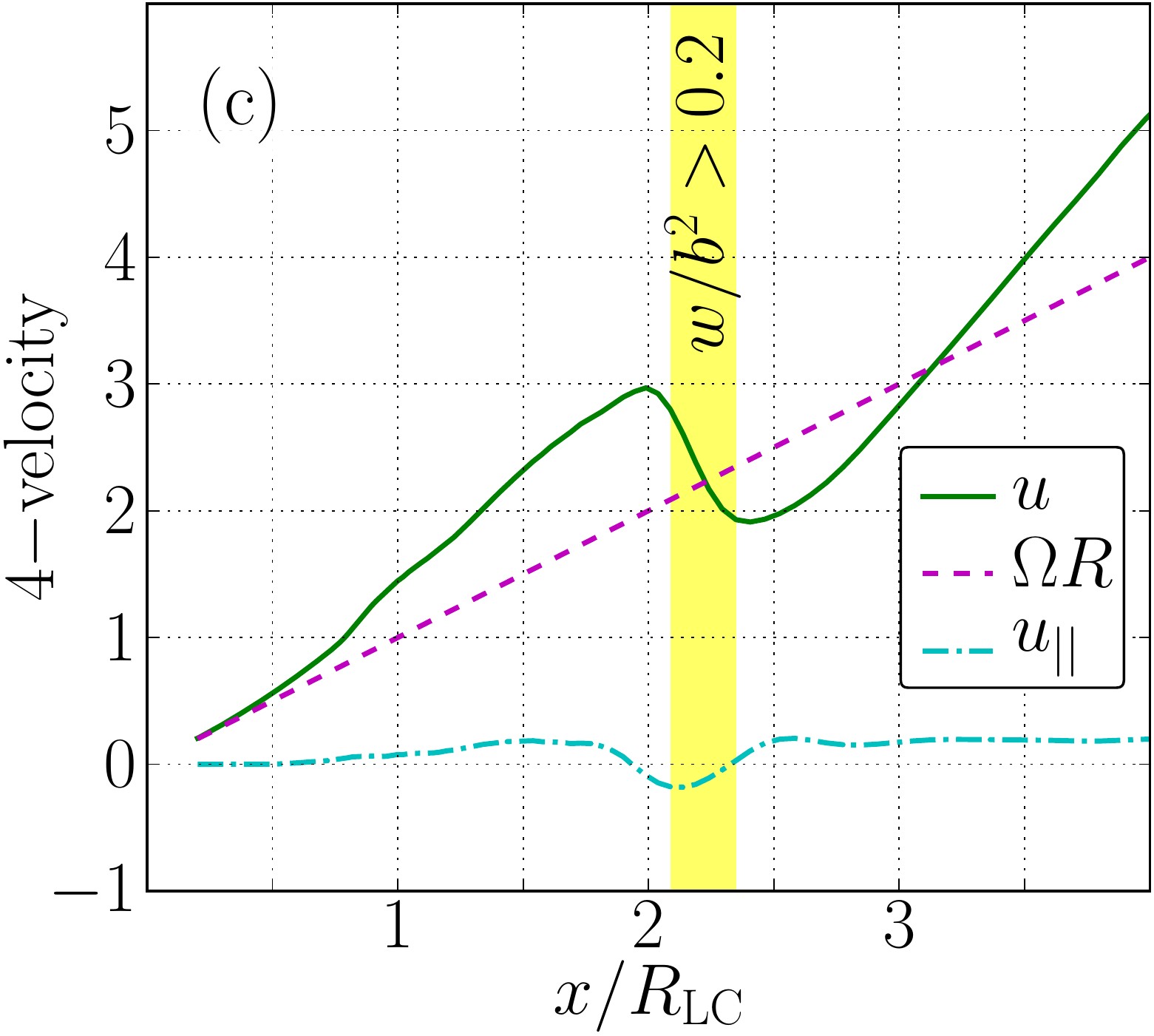}\hbox{\hspace{0.15cm}}\hbox{}\\
    \includegraphics[width=0.308\textwidth]{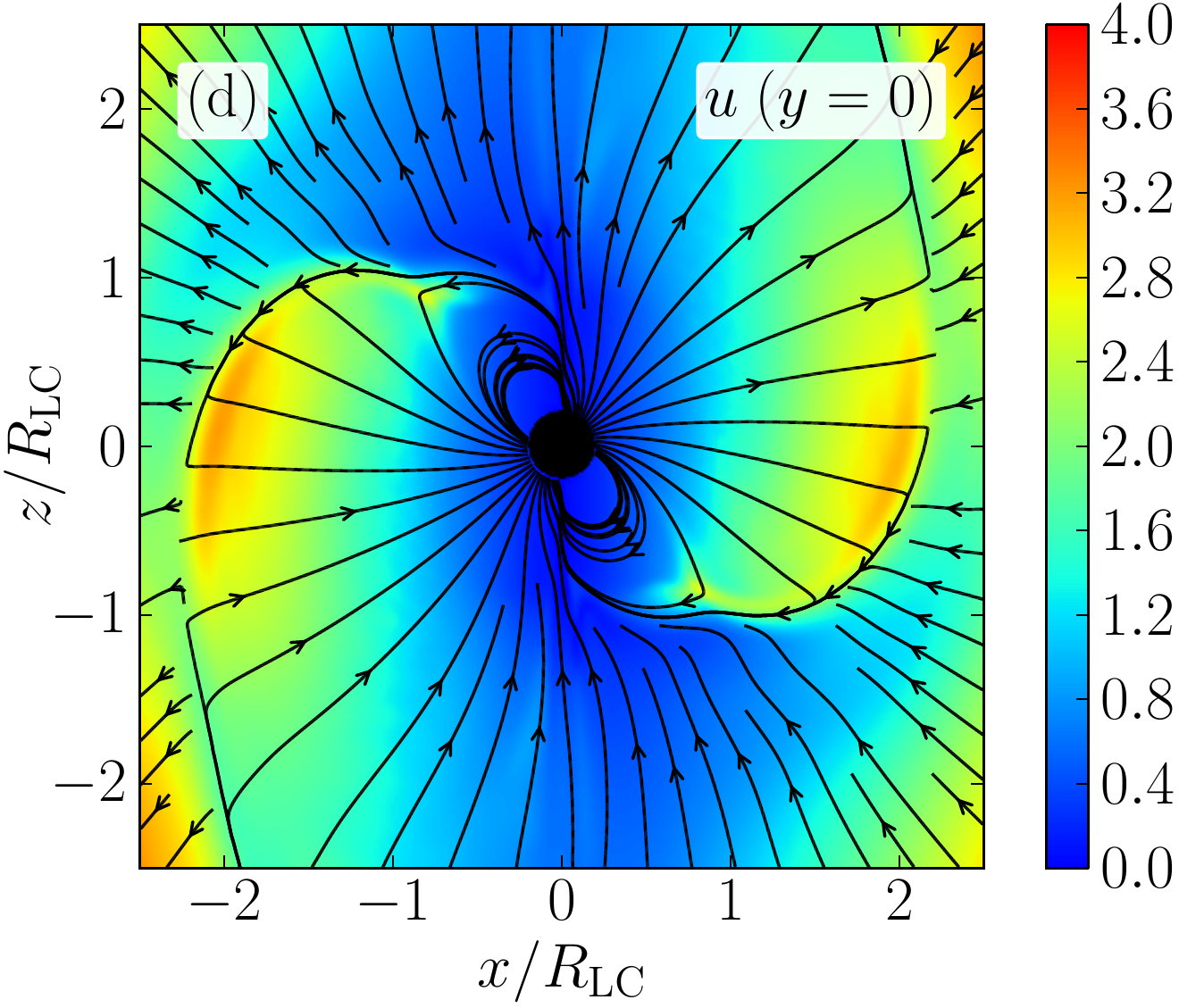}\hfill\
    \includegraphics[width=0.308\textwidth]{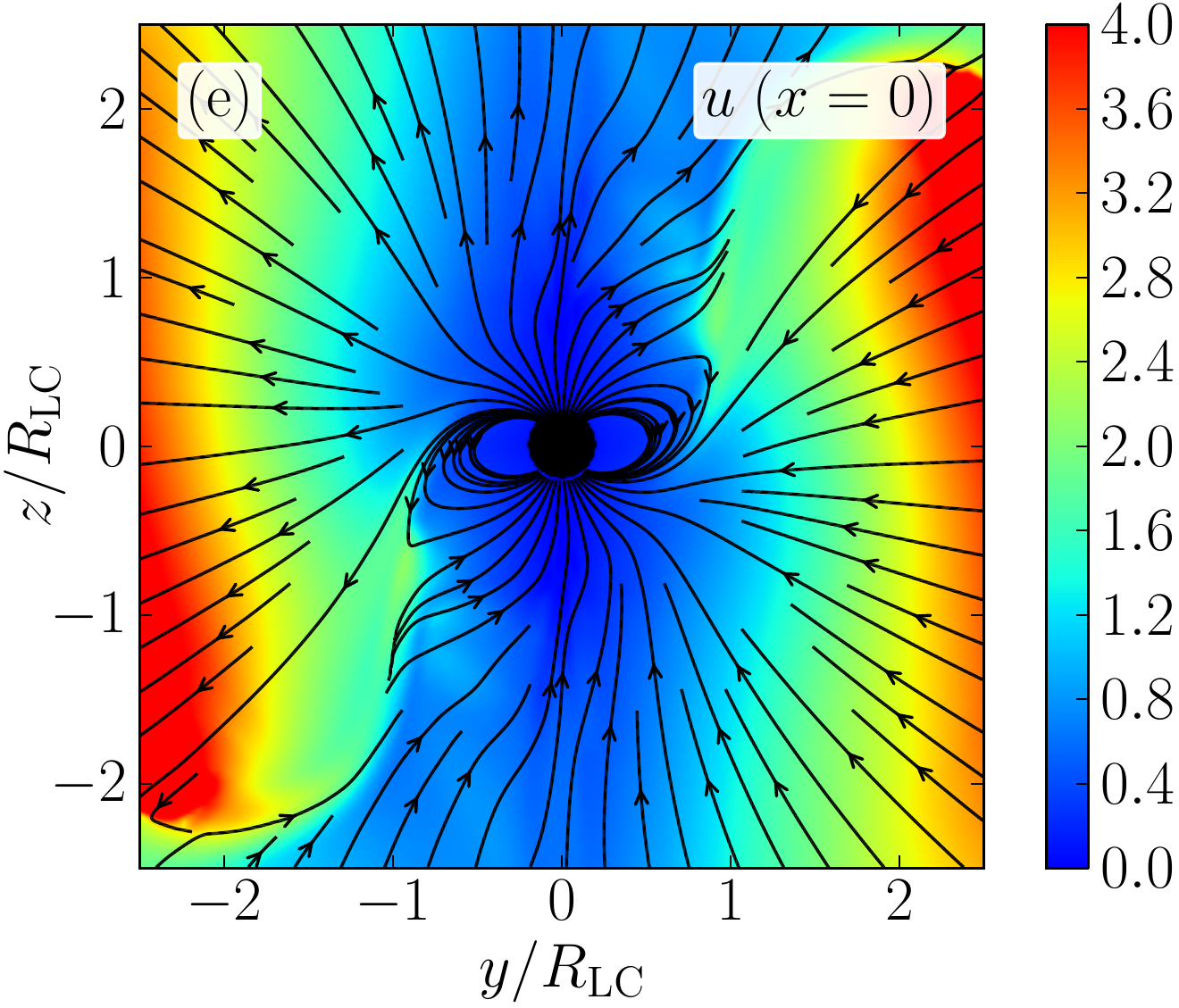}\hfill\
    \includegraphics[width=0.308\textwidth]{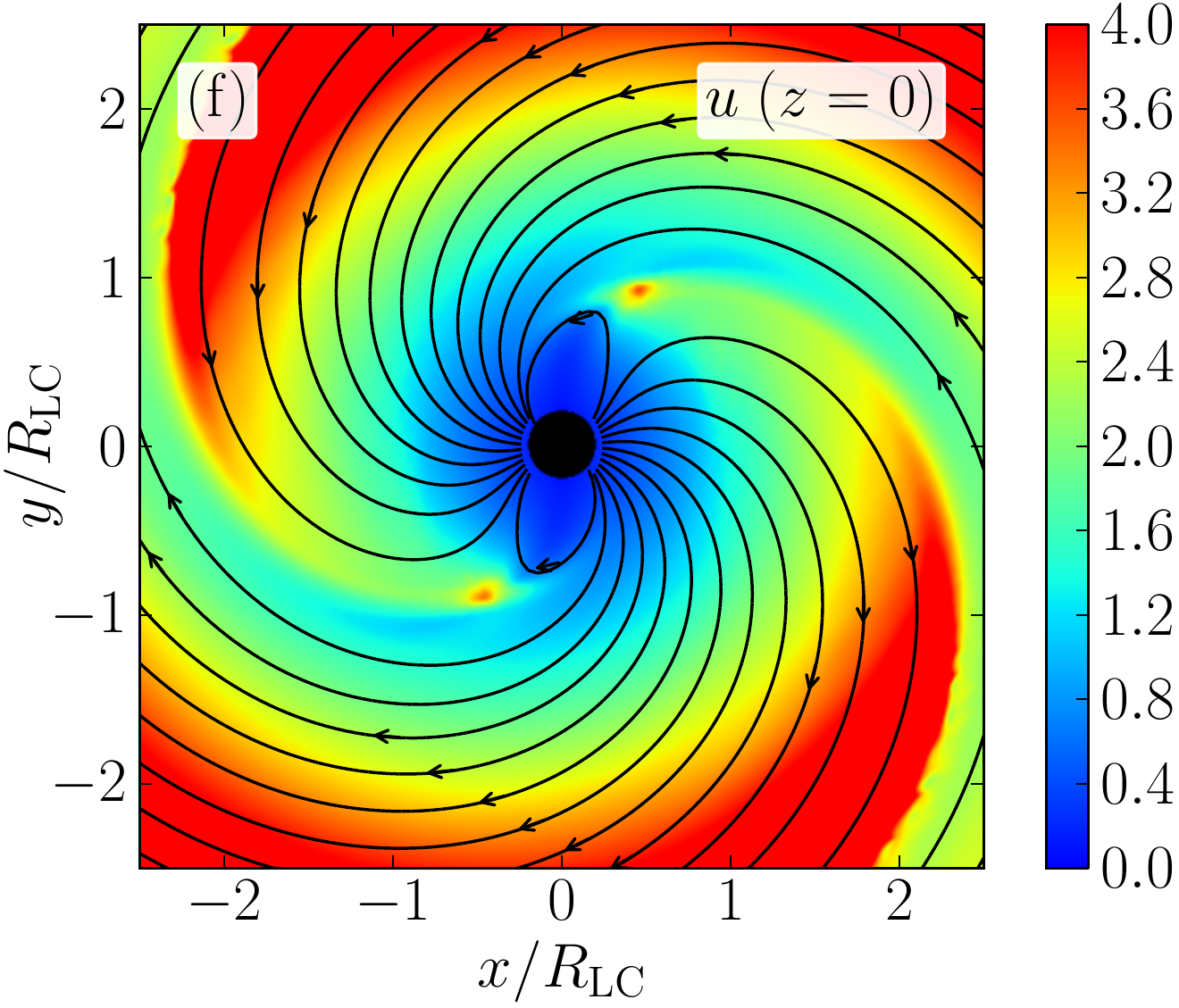}
  \end{center}
  \vspace{-0.2cm}
  \caption{Slices through a relativistic MHD
    simulation of an oblique pulsar magnetosphere ($\alpha=60^\circ$,
    model D60) taken after $3$ rotations. See
    Supporting Information for movies.  Solid lines show field
    lines as traced in the image plane. Panels (a) and (b) show slices
    in $\vec\mu{-}\vec\Omega$, or $x-z$, plane.  {\bf [Panel (a)]} Shows
    out-of-plane magnetic field component, $B_\otimes$ with colour
    (red -- into plane, blue -- out of plane).
    {\bf [Panel (b)]} Shows quantity $\log_{10}(w/b^2)$ in color. It
    is low near the star, indicating a highly magnetized flow, and high in the current sheet, indicating the
    importance of thermal pressure support.  {\bf [Panel~(c)]} Runs of
    4-velocity components vs.\ radius along the $y=z=0$ line for the
    simulation, showing the proper velocity, $u\equiv\gamma v$ (green solid), 
    and its component parallel to the magnetic field, 
    $u_{||}$ (dash-dotted cyan line), and for the analytic
    split-monopolar force-free solution the proper drift velocity, 
    $u^{\rm ff,mono}\equiv\Omega
    R$ (dashed magenta line).  
    The
    location and thickness of the current sheet, which is centered at 
    $x\approx2.2R_{\rm LC}$, is indicated by the
    yellow stripe. {\bf [Panels~(d)--(f)]} Show colour maps of $u$ 
    in $3$
    orthogonal slices (see legends).  In the current
    sheet, which is the putative source of $\gamma$-ray photons, $u$
    undergoes rapid changes that are caused by a reconnection-induced
      inflow into the sheet and that can affect
    directionality and beaming of emergent radiation (see the main text for details). 
}
\label{fig:tilted}
  \vspace{-0.5cm}
\end{figure*}
 
We carried out a number of simulations for different values of the magnetic
dipole inclination angle relative to pulsar rotation axis, $\alpha$,
from $0^\circ$ (aligned rotator) to $90^\circ$ (orthogonal rotator). 
We refer to these as models Dxx, where xx is $\alpha$ measured in degrees, see Table~\ref{tab:models}. 
We indicate resolution used in a model via a suffix ``R'' followed by the
number of grid cells in the $r-$direction; we omit the suffix for
our default choice, $N_r=256$. Our $\theta$-resolution is tied to $N_r$ via
$N_\theta=0.5N_r$, so the
aspect ratio of computational cells is about unity, 
$\Delta r:r\Delta\theta:r\Delta\varphi\approx1:1:2$. 
For all MHD models we use an ideal gas
equation of state, $p_g = (\Gamma-1) u_g$, with the polytropic index,
$\Gamma = 4/3$, appropriate for a relativistically hot
pair plasma. 

Figures~\ref{fig:aligned}(a),(b) show the structure of magnetic field and the ratio of gas enthalpy to magnetic energy in    
our
highest resolution aligned model, D0R2048, which is a 2D simulation. 
Since our relativistic MHD models are highly magnetized, with
magnetization inside the LC $\sigma_t=100\gg1$ 
(\S\ref{sec:numer-meth-probl}), they display the same generic
features as force-free models.
The currents in the magnetosphere and in the equatorial sheet cause 
the magnetosphere to open up and
form a radial Poynting-flux-dominated wind 
(e.g., \citealt{mic73}). However, not all of this wind
reaches infinity: part of it enters the current sheet and heats it,
possibly causing the sheet to  
produce high-energy emission.
Whereas force-free approximation neglects
plasma thermal pressure, Fig.~\ref{fig:aligned}(b) demonstrates that in MHD
current sheets are dominated by
the plasma pressure.

We
carried out simulations of aligned pulsar magnetospheres at different resolutions (the first $6$ models in
Tab.~\ref{tab:models}), and red lines in Fig.~\ref{fig:aligned}(c) show
their radial energy flux profiles.
  The total energy flux is essentially
independent of resolution and distance, indicating that our results
are numerically converged, and agrees to within $1\%$ with
that in other works
(\citealt{gruzinov_pulsar_2005}; \citetalias{kom06}; \citetalias{spit06}; \citealt{mck06pulff}),
\begin{equation}
  \label{eq:alignededot}
  L_{0} = \frac{\mu^2 \Omega^4}{c^3}.
\end{equation}

We quantify the amount of dissipation in the wind zone as a
fraction of total
energy flux dissipated in the interval \hbox{$r<5R_{\rm LC}$} via
$\epsfive = 1- L_{\rm EM}(5R_{\rm LC})/L(R_{\rm LC}).$
Table~\ref{tab:models} and Fig.~\ref{fig:aligned}(c) show that
$\epsfive$ monotonically decreases with increasing resolution,
$\epsfive\propto N_r^{-1/3}=(2N_\theta)^{-1/3}$: our axisymmetric
MHD magnetospheres asymptotically (in the limit of infinite
resolution) become dissipationless. 
This is to be expected: 
the level of current sheet dissipation is controlled by the
magnetospheric resistivity, which in our approach is determined by the
numerical resolution 
(see also \citealt{lm12}).

Our relativistic MHD magnetosphere is similar to the one obtained
using the low-dissipation force-free code by
\citetalias{spit06}: most of the field lines that cross the surface of LC
open up to infinity, with only a small fraction of them
entering the current sheet, where they dissipate a vanishingly small
fraction of pulsar spindown energy. Note that due to high numerical
dissipation, standard force-free codes often reach a very
\emph{different} solution: \emph{most} poloidal magnetic field lines close through the 
midplane, where they dissipate \emph{most} of pulsar spin-down
energy. 
\citetalias{kom06} noted this effect, and we do as well when we use the force-free version of \harm{}
\citep{mck06ffcode,lm12}.  Namely, we find very high levels of
dissipation, $\epsilon\sim50\%$, that do not decrease with
increasing resolution (see Tab.~\ref{tab:models}).  Such absence of
decrease is unphysical: had the models included thermal pressure
produced by current sheet dissipation, that pressure would have slowed
down the inflow into the current sheet, suppressed the reconnection,
and led to lower values of $\epsfive$
\citepalias{kom06}.  Recently,
\citet{gru08sfe,gru11sfecyl,gru11sfe50percent,gru11sfe10percent,gruz12b}
considered resistive force-free simulations of pulsar magnetospheres
and argued that the current sheet dissipates up to  $50$\%
of pulsar spin-down energy and that an increase in resolution does not
change the dissipation rate.  
Our results suggest that
the results of
\citet{gru08sfe,gru11sfecyl,gru11sfe50percent,gru11sfe10percent,gruz12b}
for axisymmetric pulsars are dominated by the large numerical resistivity of their numerical
scheme and converge to the unphysical, dissipative force-free
solution. 

Why does the force-free \harm{}
(\citealt{mck06ffcode}, Lyutikov \& McKinney \citeyear{lm12}) and many other force-free codes (e.g.,
\citetalias{kom06}; \citealt{gru11sfe10percent,gruz12b,petri12a})
show such high levels of dissipation? To handle
discontinuities in the flow, force-free \harm{} uses a Lax-Friedrichs
Riemann solver that is not specialized to treating current sheets
and hence spreads the sheet over several grid
cells. However, no force-balance inside the sheet is possible since in force-free there is no thermal pressure
to slow down reconnecting fields.  Unless one
prescribes a zero velocity of inflow into the sheet
\citep{mck06pulff}, the lack of force-balance across
the sheet in force-free causes rapid
reconnection 
\citepalias{kom06}.
In contrast, the force-free scheme
by \citetalias{spit06} can 
treat current sheets as true unresolved
discontinuities, so reconnection in the sheet is minimal.  As evidenced by our MHD results, 
we believe that in the limit of low reconnection the low dissipation force-free solutions as in \citetalias{spit06} are more representative of the physical pulsar magnetospheric shape than dissipative force-free solutions with uncontrolled numerical reconnection rate \citep[e.g.,][]{gru11sfe10percent,gruz12b}.

We now consider oblique models,
applicable to the majority of pulsars. We present results for
magnetization $\sigma_t=50$ (results at $\sigma_t=100$ are 
similar and not shown).
Figure~\ref{fig:tilted}(a) shows the $\myvec\mu{-}\myvec\Omega$ plane
for our oblique model D60:
electromagnetic quantities in our relativistic MHD models
reproduce, as expected, major features of
oblique force-free solutions of pulsar magnetospheres (see
Fig.~2(a) in \citetalias{spit06}), such as the formation of 
closed and open field line zones and the undulating equatorial current
sheet. Please see Supporting Information for movies. 
Additionally, MHD models provide crucial information about
plasma properties, e.g., velocity and
temperature in the current sheet, which are needed for light curve
computation but are missing from a force-free description. 
Although resistivi\-ty in our scheme is numerical, the
reconnection displays physical characteristics. 
Fig.~\ref{fig:tilted}(b) shows that thermal and magnetic pressures
are of the same order inside the current sheet, as in our
axisymmetric models (see
Fig.~\ref{fig:aligned}b). Hence, the thermal pressure is dynamically
important in the current sheet and affects the fluid velocity there. 

Figure~\ref{fig:tilted}(c) shows the radial dependence of velocity along the line $y=z=0$. Near the star
the proper velocity, or spatial component of $4$-velocity,
$u\equiv\gamma v$, follows the split-monopole force-free model of an oblique
rotator \citep{bog99b}, 
$u^{\rm ff,mono}= \Omega R$
\citep[e.g.,][]{nar07}.  However, further out we clearly
have $u>\Omega R$, followed by a sharp drop in $u$ across the sheet
and $u<\Omega R$ on the
other side of the sheet.
This discontinuity in $u$, which naturally emerges due to a
reconnection-induced inflow of fields and plasma into the
sheet, is neglected in the split-monopole model. 
Inside the current sheet, the velocity components in the
simulation appear to pass through the split-monopole values (e.g., $u=\Omega
R$).
This
suggests that in a volume-averaged sense the split-monopole model might
give a reasonable description of velocity in the current sheet even in the
presence of reconnection.
The
lower row of panels in Fig.~\ref{fig:tilted} shows three orthogonal
cross-sections through our fiducial oblique model, D60. 
The discontinuity in $u$ is clearly co-spatial with
the sheet. We find a similar behavior of velocity in our
force-free \harm\ models, which reaffirms that qualitatively magnetospheric structure is insensitive to the
microphysics of the model. 
At the
stellar surface we assume that the plasma is at rest relative to the
star and its proper
velocity component parallel to the magnetic field vanishes, $u_{||}=0$. Unlike force-free,
MHD allows us to compute
$u_{||}$ self-consistently  in the
bulk of the flow.  We find $u_{||}>0$, except near the current sheet, i.e.,
the plasma predominantly streams along magnetic field lines
away from the star. 
However, since $u_{||}\ll u$, \hbox{this streaming has negligible effect on plasma net velocity.}

\begin{figure}
  \begin{center}
    \includegraphics[width=0.7\columnwidth]{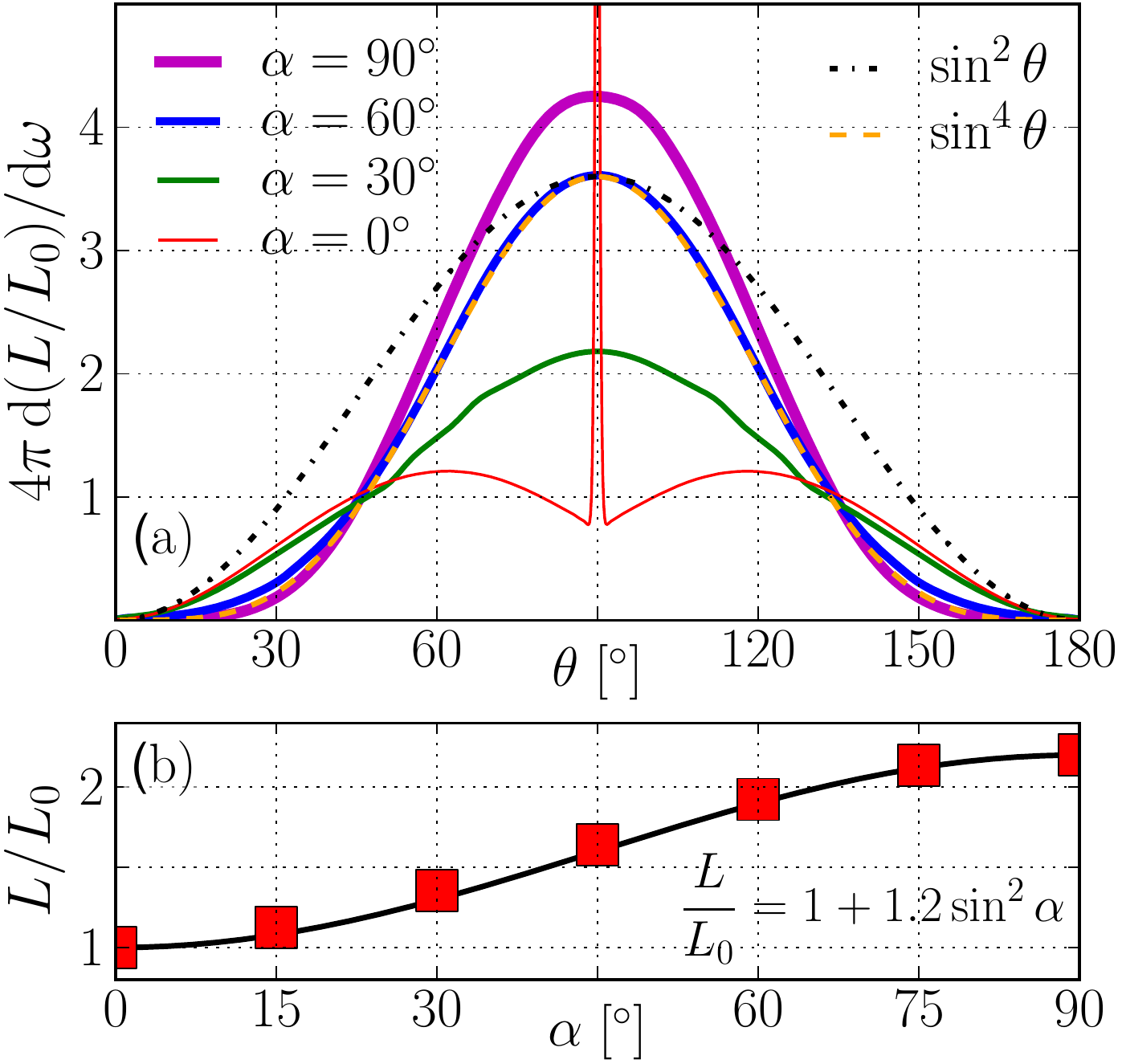}
  \end{center}
  \vspace{-0.2cm}
  \caption{{\bf [Panel (a)]} Lateral distribution of pulsar wind
    luminosity per unit solid angle, $dL/d\omega$, for models D0R2048,
    D30, D60, D90, as measured at $r=2R_{\rm LC}$. The spike
    at $\theta=90^\circ$ for D0R2048 model is due to kinetic+thermal energy outflow along
    the equatorial current sheet. 
    At high ($\alpha\gtrsim60^\circ$) obliquity
    pulsar wind luminosity is more equatorially
    concentrated (dashed line) than in the split-monopole wind model
    (dash-dotted line).
    {\bf [Panel (b)]} Pulsar luminosity increases with increasing
    obliquity angle, $\alpha$, for our MHD models, D0--D90, 
    in excellent agreement with \citetalias{spit06}.}
\label{fig:edot}
  \vspace{-0.5cm}
\end{figure}
 
Figure~\ref{fig:aligned}(c) shows the radial profiles of total and Poynting
(electromagnetic) energy fluxes for models at different values of
pulsar obliquity.  The
total energy flux, $L(r)$, is conserved to better than $3\%$ accuracy in all
cases.
We confirmed that our oblique models are numerically converged: 
an $8\times$ increase
in resolution affects the spindown rate by $\le2\%$ (compare
models D60R64, D60R128, D60 and D60R512 in
Tab.~\ref{tab:models}). 
Based on this, we conservatively estimate that the accuracy of our spin down energy loss
measurements is $\simeq5\%$. 
Interestingly, Table~\ref{tab:models} shows that changing resolution
for our tilted model D60 
leads to much smaller changes in the amount of
magnetospheric dissipation, $\epsfive$, than in our axisymmetric
models. This suggests that the 3D motion of the current sheet through the
numerical grid increases the dissipation in our scheme.

Table~\ref{tab:models} shows that the amount of
magnetospheric dissipation in our force-free \harm{} models, D0ff-D90ff, dramatically
decreases as inclination angle increases. Namely, dissipation is
unacceptably high, $\epsilon\gtrsim25\%$, at low inclination angles,
$\alpha\lesssim30^\circ$. However, the dissipation becomes much 
smaller, $\epsilon\lesssim 5\%$,
at higher inclination angles, $\alpha\gtrsim60^\circ$. 
This low level of
magnetospheric dissipation (in fact, even smaller than in our MHD
models) for highly inclined force-free \harm{} pulsars 
indicates that force-free \harm{} models D60ff and D90ff provide a
good description of the electromagnetic part of pulsar magnetosphere. 
Our MHD models do
not show such strong trends of $\epsilon$ vs.\ $\alpha$ and are 
applicable at all inclination angles. It is likely that the improved dissipation properties of force-free schemes at large inclinations have to do with the larger fraction of the current in the sheet that is carried by the displacement current for higher obliquity. Force-free schemes without conduction currents become vacuum-like in the sheet region, and this may suppress reconnection there. This is not the case for MHD schemes which still have to include the plasma in the current sheet.

It was suggested that aligned
\citep{1973ApJ...186..625I,1974ApJ...187..585M} and oblique
\citep{bog99b} pulsar magnetospheres resemble the split-monopole wind
asymptotically far from the star. That the field lines in
Figs.{}~\ref{fig:aligned} and \ref{fig:tilted} are predominantly
radial supports this suggestion.  However, \emph{quantitatively},
pulsar wind substantially differs from the split-monopole.  As noted above,
reconnection-induced inflow into the magnetospheric current sheet
modifies the velocity of the wind.  The lateral distribution of wind
luminosity flux also deviates from split-monopole's
$dL/d\omega\propto\sin^2\theta$.  For the aligned pulsar, instead of
peaking at the equator, the wind luminosity is double-peaked (red line
in Fig.~\ref{fig:edot}a).  For highly inclined pulsars, with
$\alpha\gtrsim60^\circ$, the wind luminosity
is well-described by $dL/d\omega\propto\sin^4\theta$ (solid blue line
and orange dashed line in Fig.~\ref{fig:edot}a are on top of each
other) and is thus substantially more equatorially-concentrated than
the analytic split-monopole expectation,
$dL/d\omega\propto\sin^2\theta$ (dot-dashed black line in
Fig.~\ref{fig:edot}b).  This has potentially important consequences
for the theoretical modeling of pulsar wind nebulae, where the angular
\hbox{distribution of wind luminosity can be directly observed}. We confirmed the 
deviations from split-monopole using MHD and force-free versions of \harm{}, and
with force-free code of \citetalias{spit06}.  

Figure~\ref{fig:edot}(b) shows that pulsars at high obliquity lose
larger amounts of energy than at low obliquity.  Both in our MHD and
force-free \harm{} models the spin-down power, $L$, is well-described
by the analytic fitting formula, $L/L_{0} = k_1(1+k_2\sin^2\alpha$),
in good agreement with force-free results of \citetalias{spit06},
albeit with slightly different values of numerical factors, $k_1 = 1$
and $k_2 = 1.2$.

\section{Conclusions}
\label{sec:conclusions}

We obtained axisymmetric and oblique pulsar magnetosphere solutions 
using time-dependent relativistic MHD equations in 3D.
We used a conservative relativistic MHD formulation that allowed us to
account for resistive heating and thermal pressure support in
magnetospheric current sheets, both of which are important for
obtaining numerically converged solutions (see \S\ref{sec:results}).
Our solutions are highly magnetized, with $\sigma = b^2/4\pi\rho c^2
\simeq 50{-}100$ near the LC, and are, therefore, close to 
force-free.  We verified that the electromagnetic spin down power in our
relativistic MHD models quantitatively agrees with force-free models.
Our MHD models generalize force-free solutions by providing crucial
information about fluid motions that is missing from a force-free
description: plasma density, pressure, and the velocity component
parallel to the magnetic field, $u_{||}$. Knowing this
information is required for computing the beaming and phase of
current sheet's $\gamma$-ray emission 
\citep{bs09b,bs09a} either due to thermal or
non-thermal particles \citep{2012arXiv1208.2819A,2012arXiv1210.3346U}.
These calculations will be presented in an upcoming publication. 
We note that while MHD models provide a complete
description of plasma motion along the open field lines, we still have to
switch to a force-free-like description inside the LC to avoid mass
and internal energy build-up on the closed field lines
(see \S\ref{sec:numer-meth-probl}).

We showed that the conventional expectation that the magnetospheric
structure is well-described by the split-monopole wind model does not hold
\emph{quantitatively}: at high inclinations, $\alpha\gtrsim60^\circ$, 
the pulsar wind luminosity is more equatorially concentrated than in a
split-monopole wind and the wind velocity structure is modified by
reconnection-induced inflow into the magnetospheric current sheet.  Our work considered a non-relativistic outflow from the surface of the
NS ($u_{||}=0$). We plan to investigate if an ultra-relativistic
outflow from the surface ($u_{||}\gg1$) can cause ultra-relativistic velocity
inside the current sheet on the scales
of LC. 
As magnetospheric conductivity can vary, possibly influenced
by the amount of magnetospheric plasma supply \citep{lst12,lst11}, resistive
relativistic MHD codes should be developed \citep{komi07resistive,palenzuela2009,dio12}
and used to study physical resistivity effects on the structure
of magnetospheric current sheets and $\gamma$-ray light curves
\citep{lst12,lst11,kalap12b,kalap12a}. This will also allow studies of
plasma accumulation and plasmoid formation near the Y-point
that can explain pulsar glitches and associated
changes in pulsar braking indices (\citealt{cont05,buc06}; \citetalias{spit06}).

\section*{Acknowledgments}
AT is supported by the Princeton Center for Theoretical Science
Fellowship. AS is supported by NSF
grant AST-0807381 and NASA grants NNX09AT95G, NNX10A039G,
NNX12AD01G. We thank the anonymous referee for
insightful suggestions that helped improve the manuscript and Jonathan
C.\ McKinney and Alexander
Philippov for fruitful discussions. The
simulations presented in this article used computational resources
supported by the PICSciE-OIT High Performance Computing Center and
Visualization Laboratory, and by XSEDE allocation TG-AST100040 on 
NICS Kraken and Nautilus and TACC Ranch. 

\section*{Supporting Information}
\label{sec:si}
Additional Supporting Information is available for this article.\hfill\newline
{\bf Movie files.} Movies of model D60: movie
\href{http://youtu.be/cjua6XhhNL4}{1 (link)} and
\href{http://youtu.be/dUR2Lx1JGRM}{2 (link)}.


\label{lastpage}
\end{document}